\documentclass[runningheads]{llncs}
\usepackage[utf8]{inputenc}
\usepackage[]{url}

\usepackage{graphicx} 
\usepackage{color} 

\usepackage{array}
\newcolumntype{L}[1]{>{\raggedright\let\newline\\\arraybackslash\hspace{0pt}}m{#1}}
\newcolumntype{C}[1]{>{\centering\let\newline\\\arraybackslash\hspace{0pt}}m{#1}}
\newcolumntype{R}[1]{>{\raggedleft\let\newline\\\arraybackslash\hspace{0pt}}m{#1}}

\usepackage{float} 
\usepackage{multirow} 
\usepackage{subcaption} 
\usepackage{csquotes} 
\usepackage{longtable} 
\usepackage{todonotes}

\begin{document}
\title{Intelligo ut Confido: Understanding, Trust and User Experience in Verifiable Receipt-Free E-Voting (long version)}
\titlerunning{Intelligo ut Confido}
\authorrunning{Zollinger, R\o{}nne, Schneider, Ryan and Jamroga}

\author{Marie-Laure Zollinger\inst{1}, Peter B. R\o{}nne\inst{1}, Steve Schneider\inst{2}, \\ Peter Y. A. Ryan\inst{1} and Wojtek Jamroga\inst{3}}
 
\institute{Interdisciplinary Centre for Security, Reliability and Trust, University of Luxembourg \and Surrey Centre for Cyber Security, University of Surrey, UK
\and
Institute of Computer Science, Polish Academy
of Sciences}
 

\maketitle

\begin{abstract}
Voting protocols seek to provide integrity and vote privacy in elections. To achieve integrity, procedures have been proposed allowing voters to verify their vote -- however this impacts both the user experience and privacy. Especially, vote verification can lead to vote-buying or coercion, if an attacker can obtain documentation, i.e. a receipt, of the cast vote. Thus, some voting protocols go further and provide mechanisms to prevent such receipts.  
To be effective, this so-called \emph{receipt-freeness} depends on voters being able to understand and use these mechanisms. 
In this paper, we present a study with 300 participants which aims to evaluate the voters' experience of the receipt-freeness procedures in the e-voting protocol Selene in the context of vote-buying. This actually constitutes the first user study dealing with vote-buying in e-voting. While the usability and trust factors were rated low in the experiments, we found a positive correlation between trust and understanding.

\end{abstract}


\section{Introduction} \label{understanding:sec:intro}

Voting and elections are a prime example of multi-agent systems~\cite{Shoham09MAS}, where humans interact in a carefully designed technological environment~\cite{Bella14concertina}. This applies even more obviously to electronic voting~\cite{Hao17evoting}.
To this end, \emph{voting protocols} are designed to satisfy certain important properties, in particular Privacy and Integrity. Privacy is often defined by three sub-properties, namely Ballot-Secrecy, Receipt-Freeness and Coercion-Resistance. Ballot-Secrecy ensures that the protocol does not reveal the voter's choice. Receipt-Freeness says that the system will not provide any evidence that allows a voter to prove how they voted, e.g. to prevent vote-buying. Finally, Coercion-Resistance is closely related to Receipt-Freeness and allows the voter to pretend to cooperate actively with a coercer~\cite{Delaune06coercionres}, but still cast the intended vote. The important difference is that when interacting with a vote-buyer, a voter has an economical incentive trying to obtain a receipt of the vote. 
A vote buyer, as defined in the security literature of e-voting, is offering a voter money for a vote cast according to the buyer's preferred choice, but the money is only paid upon receiving a receipt for the corresponding vote. However, if the receipt can be faked by the voter, the vote buyer cannot trust the receipt and hence vote buying should be disincentivised.

Integrity means that the announced outcome of the election is correct. 
Typically we demand more: the system should also deliver a proof that the result is correct. 
This can be ensured by means of \emph{end-to-end verifiable voting protocols}~\cite{Ryan15verifiability}. This entails two complementary verification procedures: firstly, universal verifiability means that anyone can check that the vote count is correctly established from the submitted electronic ballots, secondly, individual verifiability means that each individual voter can check that their vote intent was correctly captured. The latter is most interesting from a user perspective since it inherently needs user interaction. 
For instance, in the Selene e-voting protocol~\cite{DBLP:conf/fc/RyanRI16} all voters receive a tracking number which points to their plaintext vote in the election outcome. 
They can also choose a fake tracking number to show to e.g. a vote-buyer, providing to the voter a \textit{receipt-free mechanism}. A vote buyer cannot determine whether the voter presents a real or fake tracker, and hence does not have proof of how the voter actually the voted. 
The voter's understanding and the user experience of the verifiability procedures in Selene were explored in several papers~\cite{verena,zollinger,2021marky,2021zollinger}. However those studies did not include the receipt-free 
mechanism which can introduce trust issues and misconceptions about the protocol.

Receipt-Free or Coercion-Resistance mechanisms have rarely been tested with end-users; to our knowledge, only \cite{DBLP:conf/ihc/NetoLAMNT18} explored a Coercion-Resistance mechanism for the JCJ e-voting protocol \cite{DBLP:conf/wpes/JuelsCJ05}, and nobody investigated Receipt-Freeness in the context of Vote-Buying. 
This is an important gap in the assessment of practical security of voting procedures, as the Privacy 
of protocols are based on the correct use of fraud prevention mechanisms. For an overview, see \cite{kulykevoteid2020}.

In this paper, we present the first large scale study of receipt-free mechanism in the Selene voting protocol. The study is based on experiments with 300 human participants recruited through the platform Prolific. We evaluated the user experience (UX), trust, and understanding of the voting procedure, and formulated three hypotheses to be tested: 

\begin{description}
\item[H1] The voting application and its receipt-free 
feature provide a positive user experience to the participants. 
\item[H2] The application and  receipt-free 
mechanism are trusted by the participants.
\item[H3] Participants who understand the receipt-free 
mechanism have increased trust in the application.
\end{description}

To evaluate the UX, we use the user experience questionnaire (UEQ).
To evaluate trust, at the time of the user experiment there was no standard questionnaire to assess this metric in the voting context. Therefore, we defined trust for voting and proposed a new questionnaire assessing the voters' trust in the protocol in section \ref{understanding:sec:trust}, especially including questions related to the vote verification and elections (quantitative analysis in Sec. \ref{understanding:sec:study}).
The correct understanding of the receipt-free 
mechanism is evaluated through the steps that the participants perform. To evaluate the understanding, we designed a user protocol with a specific scenario inspired by game theory, as experimented in \cite{DBLP:conf/uss/LlewellynSXCHRS13} for privacy in voting. A correct understanding of the mechanism will lead to a specific workflow. These elements are detailed in section \ref{understanding:sec:design}.

Finally, participants were invited to tell us why they made their choice in the game, and how they felt. We categorized their answers in a qualitative analysis (App. \ref{app:qualititative}) and correlated this with the participants' understanding (Sec. \ref{sec:rela}).

To summarize, our contributions are:
\begin{itemize}
    \item A new questionnaire to evaluate trust in the context of voting,
    \item A unique game design to assess the voters' understanding of a system, 
    \item An evaluation of the relationship between understanding and trust,
    \item A qualitative analysis of user feedback on receipt-freeness and vote-buying, 
    \item A list of recommendations for future voting systems and user studies (Sec. \ref{sec:rec}).
\end{itemize}

\section{Related Work} \label{understanding:sec:rw}
 
Our experiment is inspired by 
\cite{DBLP:conf/uss/LlewellynSXCHRS13} where a game approach was used to evaluate the understanding of the privacy mechanisms in the e-voting protocol Prêt-à-Voter (PaV) \cite{pav} (without interaction with a vote-buyer). 
In PaV voters get receipts, but it only contains an encrypted form of their vote choice and hence preserve ballot-privacy. 
They designed a simple game that they ran with 
12 participants. 
In the game, the participants tried to guess each other's votes and had the choice between publishing their receipt or not. They were given a reward of \pounds1 by revealing it, and otherwise nothing. Hence, participants who understood the system should choose to reveal the receipt as the most profitable strategy. Thus understanding could be measured, but with only 12 participants a conclusion was hard to draw. We improve on this with a large number of participants and direct interaction with a vote buyer. 



Until now, most of the studies focused on the usability and appreciation of voters for a given system, but a true evaluation of their understanding is rarely performed.
Also, it has been only evaluated in reference to predefined mental models of the participants.
In \cite{DBLP:conf/hci/AcemyanKBW15}, the authors let voters draw their mental models for three voting schemes. This study reveals that voters focused much more on the voting phase in all three protocols, as the verification features remained unclear to them. In the particular case of Selene, two studies have looked at mental models of participants \cite{zollinger,2021zollinger}. It appears that the understanding of verification was better when the participants have seen a possible threat, e.g. a vote manipulation \cite{2021zollinger}.
The verification mechanisms of Selene have also been implemented without the receipt-free 
mechanism with a commercial partner in United Kingdom \cite{sallal2019vmv}, augmenting an existing voting system. The user experience was evaluated \cite{alsadi2020} showing satisfaction and a higher confidence in the system.

The evaluation of coercion-mitigation features have rarely been performed, except for the protocol JCJ \cite{DBLP:conf/wpes/JuelsCJ05} in \cite{DBLP:conf/ihc/NetoLAMNT18}. The results revealed some usability issues that could play a role in real elections.
\section{The Selene protocol} \label{sec:protocol}

Selene is an e-voting protocol that has been designed to make the individual verification more usable and intuitive for voters. Usually, verification procedures can be categorized into four types, as described in \cite{2021marky}: audit-or-cast, verification device, code sheets and tracker data. Selene belongs to the last category, the other categories usually require the voters to either manipulate ciphertexts, or to verify codes rather than plaintext votes. 
Tracker based protocols allow voters to verify the presence of their vote in plaintext in the final tally using a (private, deniable) tracking number. The special feature of Selene is that this tracker is only delivered to the voter \textit{after} the tally is published to allow the coercion evasion strategy described below. 


\subsection{The protocol and voter experience}

The complete description of the protocol is available in the original paper \cite{DBLP:conf/fc/RyanRI16}. 
%
%
Each voter has a pair of public and private keys that are used in the verification phase. The election keys are also generated, the election public key is distributed to voters. A public bulletin board ($BB$) is used to display all public data.

\textit{1) Setup.} The election authorities generate the list of tracking numbers. The trackers are encrypted with the election public key, then shuffled and associated with the voters. A trapdoor commitment to each tracker is created and published on the bulletin board, sealing the relation between a tracker and a voter. To open a commitment and see the tracker, one needs the voter's private key and a secret (dual key) which is revealed later to the voter by the authorities.

\textit{2) Voting.} When the setup phase is over, voters can cast a vote encrypted with the election public key. The encrypted vote is published on $BB$. 

\textit{3) Tally.} After voting, the authorities retrieve the pairs of encrypted tracking numbers and encrypted votes, shuffle the pairs and decrypt them to obtain and publish the pairs of plaintext tracking numbers and votes. 

\textit{4a) Verifying.} Finally, the secret dual key associated to each commitment during the setup phase is delivered to each voter. By combining the dual key, the commitment and their private key, each voter can retrieve the tracking number, and verify the associated plaintext vote.

\textit{4b) Faking.} If a voter is interacting with a vote-buyer or being coerced
, the voter can choose an alternative tracker, showing a plaintext vote that corresponds to the adversary's request. From this tracker and the commitment, a fake dual key is computed by the voter using her private key. This can be done after the tally phase. 
The combination of this fake dual key with the commitment and private key of the voter will open to the selected fake tracking number.

For the purposes of the trial, voters could verify their own vote and later request that an alternative tracker be displayed to mislead the vote-buyer. 
Arguably, this results in a more complex experience compared to what voters would encounter in normal elections.

\subsection{Web application}

For the experiment, we implemented a web app reflecting the user steps described above.  The voter can access the following pages through a menu, after login:

\begin{itemize}
    \item \textit{Home:} this page explains the purpose of the web app and the different pages.
    \item \textit{Voting:} the voting question is displayed with the possible vote choices. 
    \item \textit{Verification:} this page allows to access the public election result as vote/tracker pairs. The voter can retrieve the tracking number to verify the vote, or choose a fake tracking number.
    \item \textit{About:} information about Selene and its features is displayed here.
    \item \textit{Contact:} a link to our email is provided in case of questions.
    \item \textit{Logout:} used to log out from the study.
\end{itemize}
A default workflow is proposed once the voter is connected.
In the voting section, after selecting the candidate, a confirmation page is displayed. If the voter decides to confirm, the voter can click on a button ``Encrypt and send my vote''. As shown in \cite{2021zollinger,2021marky}, such an  interaction  is 
seen by voters and does not require any additional skill, while increasing the security perception.
On the verification page, the general results are displayed and the voter is offered two choices: fake the tracking number in case of coercion or vote-buying, or go for verification directly.
To fake the tracking number, a new page is displayed where the voter can access the bulletin board and type the chosen tracking number. The voter is warned that it is not possible to retrieve the real tracking number after this request. After validating, the voter is redirected to the main verification page.
If the voter chooses to verify
, the app computes the (real or fake) tracking number 
and the voter can connect to the bulletin board to verify the vote.

In previous applications for Selene \cite{zollinger,2021zollinger,verena,2021marky}, the authors made the choice of highlighting the tracking number and corresponding vote directly in the application to increase usability, with the risk of lowering privacy and the security perception. In this version, we provide the tracking number and the user has to display the bulletin board and look for the tracker to verify the vote. This choice has been made to be more accurate with the original protocol design, however this also results in a less convenient interaction for the user.

\section{Trust} \label{understanding:sec:trust}

Trust is a topic that is omnipresent in studies about voting \cite{chiang2009trust,5460386,DBLP:conf/re/SchneiderLCHSX11,DBLP:journals/ieeesp/KulykNBV17,marky2018what,zollinger,agbesi2022will}. It is rather complex to evaluate, as trust can be impacted by many aspects related to elections or to the media used: trust in politics, trust in internet technologies (in the case of internet voting), understanding of the app, etc.

Furthermore, there is no standard questionnaire available to evaluate trust of users for voting systems. The UEQ+ questionnaire \cite{ueq+handbook} only proposes a few items linked to trust. To close this gap and to understand the relation between understanding and trust, we designed a more specific questionnaire for the e-voting context, especially we have questions related to vote verification and elections. We now discuss trust and the design of the questionnaire. We note that after our experiment was carried out another trust measure for voting was suggested \cite{acemyan2022trust}. However, with 44 questions it would not have been suitable for our online experiment where users have limited patience.

In \cite{luhmann}, Luhmann differentiates trust and confidence.
Confidence can be obtained without any additional explanation, in particular security does not need to be perceived to be acknowledged while trust requires an evaluation from the users' of their security perception to be granted. 

In \cite{Pieters_2010}, Pieters highlights that a voting system can obtain the voters' \textit{confidence} if it works correctly. A system that guarantees a correct result will not worry the voters. But, when a new system implementing new procedures, such as verifiability features, comes in with a comparison to the old system which has the confidence of voters, \textit{trust} and \textit{distrust} takes the place of confidence.
The author also mentions the relationship between trust and explanation.
We need the voters to \textit{understand} verifiability to implement a new system and convince them to use it.
Previous works have already mentioned the relationship between trust and the explanations \cite{DBLP:conf/iui/GlassMW08}, in particular in voting \cite{zollinger,2021zollinger}.

In this study, we aim to provide a reasonable amount of information regarding our protocol, in order to increase our chances of having a good trust rating. However, the participants will have a limited amount of time to evaluate our app, so we should not provide too much information that could overwhelm them.
 
\subsection{Our metric}

Our 
voting-oriented trust questionnaire contains 
eight questions. From the studies and literature cited above, we see that trust depends on a positive evaluation of the security. In our questionnaire, we evaluate the feeling of security on one hand; and the acceptance of the system on the other, to see if trust is engendered. The questions labelled by these topic are 
1) [Acceptance] ``I trust the system and I would use it in a real election''. 2) [Security] ``I believe that the personal information (vote included) is kept private''. 3) [Security] ``I think that the system ensures the integrity of the elections''. 4) [Security] ``I think that the system is transparent and lets me know everything about its behaviour''. 5) [Acceptance] ``I think that the verification phase is important''. 6) [Security] ``I was convinced by the verification phase that my vote was correctly recorded''. 7) [Acceptance] ``I would use such a verification system if it was available''. 8) [Security] ``I think that the result of the election can be changed by an attacker''. Answers were given on a Likert scale with 6 choices from strongly disagree to strongly agree. The result was linearly transformed so that each question could give 0-10 points, with 10 indicating maximal trust. 
%
%
%
To label the outcomes, we used the following classification: High trust for a score $>64$, moderate trust for $48-63$, 
low trust for $32-47$ and
very low trust for a score $<32$.

\section{User protocol} \label{understanding:sec:design}

For the experiment we used the crowd-sourcing platform Prolific \cite{prolific}. 
\subsection{Methodology}
The context provided to the participants was the following: the city council is organising local elections to request its citizens' opinion on several society subjects. To cast their vote, the participants 
used our online application. 

Trust and UX were evaluated in the standard way: after having interacted with the application, the participants were given questionnaires. We used the System Usability Scale, the User Experience questionnaire \cite{ueqhandbook} and our trust questionnaire. 
Then, to evaluate the understanding, we designed a user game inspired by the game theoretic experiment in \cite{DBLP:conf/uss/LlewellynSXCHRS13}. The participant had to interact once more with the application but we provided an additional scenario: the participant had to interact with a vote buyer\footnote{With Selene, countering vote buying and 
coercion involves the same user steps.}. The instructions from the vote buyer were displayed in a box next to the web page: the vote-buyer asks for a different vote than the choice made by the voter (we configured the game by asking in advance the voter's opinion, see below). Our evaluation consists in looking at the participant's behaviour in such a  scenario. Our assumption is that a correct understanding will lead the participants to vote for their candidate and use the receipt-free 
mechanism to provide a fake tracker to the vote buyer. 

\noindent\textit{Pilot studies.} We ran two pilot studies with five participants in each. 
In the first pilot, none of the five participants watched the video nor tried the receipt-free 
mechanism (even with the vote-buying scenario) and they finished the study in less than five minutes (while 20 min.  
were given). This rush bias is well known and called ``satisficing'' in Prolific's  terms of use. To ensure the participants use the app fully, we introduced a workflow: they could not access the questionnaires and continue the study before they used the mechanism to get a new tracking number. Guidance was provided as side notes on the website. 
Also, some attention checks were added to the questionnaires as recommended by Prolific. We further discuss the limitations in section  \ref{method:subsec:limitations} below.


Participants were paid 2.5\pounds{} for the study (20 minutes) which was evaluated as a \textit{Good} hourly rate by Prolific, and we added an extra 1\pounds{} as a bonus payment for having played the game.

After a consent form, the user experiment had the following steps


\textbf{Demographics} We recruited 300 participants on the crowd-sourcing website Prolific \cite{prolific}. We used the pre-screening feature to select participants: to ensure that they have a similar experience in voting, we chose UK citizens living in UK. The average age was 33 years (Min=18, Max=73, SD=11). They come from various backgrounds, the education level differed: No diploma (0,67\%), A-Levels (13,33\%), College Level (19,33\%), Bachelor (42,33\%), Master Degree (20\%), PhD (1,33\%) and other (3\%). Finally, regarding their attitude toward online voting, 2,33\% were negative, 7\% were rather negative, 39,67\% were neutral, 35,67\% were rather positive and 14,67\% were positive.


\textbf{Configuration} In the end of the demographics' questionnaire, we asked the participants to answer the voting question used in the game, to configure the vote buyer's instruction. The question was about the COVID-19 crisis:

\noindent\rule{\textwidth}{0.4pt}
Regarding the recent events related to the COVID-19 pandemic, according to you, what would be the best policy to adopt at the beginning of the epidemic?\\
- A strict confinement for all\\
- No confinement but detection tests available for everyone\\
\noindent\rule{\textwidth}{0.4pt}
We had no interest in the answer, we configured the game by changing the vote buyer's instructions according to their opinion. If they chose ``A strict opinion for all'', the vote buyer asks for ``No confinement but detection tests available for everyone'' and vice versa.

\textbf{Video} explaining the protocol: We describe the Selene protocol in a 4-minute video that the voter was invited to watch.


\textbf{A tutorial to demonstrate the receipt-free mechanism} First, we let the participants use the application through a tutorial. As mentioned above, the first pilot study has shown that participants were rushing to end the study as fast as possible. The tutorial ensures that they see and test all available features in the application, a specific workflow was forced with guidance, given as side notes.  Therefore, participants were able to verify their vote and then fake their tracking number. We wanted to show that they can see their plaintext vote, but also have the ability to change their tracking number to show another vote to a coercer or vote-buyer.

\textbf{Questionnaires} We evaluated the usability, user experience and trust after this tutorial phase. The reason was that we did not want to influence their trust rating by going through a coercive scenario, but obtain their general impression of the app. Also, we put a few attention checks (through questions about the app) at the beginning of our questionnaire. The checks were announced in the study description on Prolific. Our goal was to increase the attention given to the explanations in the app. 
Of course there is a possibility that the participant did not understand the protocol and provide wrong answers. We did not exclude such participants, our goal was to help them to focus on the information rather than skipping it as in the pilot study.



\textbf{Vote-Buying Game} We introduced the game by telling the participants that they will receive instructions from a vote buyer. The rules were given as follows:

\noindent\rule{\textwidth}{0.4pt}
A vote buyer wants to buy your vote by giving you a vote instruction. He may ask you how you voted and to reveal your tracking code, in which case you can give an alternate code.\\
If you send a tracking code for the requested candidate, you will receive 70 pence from the vote buyer.\\
If you want to keep your vote intention, you will receive 30 pence.\\
\textbf{These incentives will be provided as bonus payment after the study.}\footnote{In the end we provided both incentives as bonus payment to all participants regardless of their choice, for fairness. }\\
\noindent\rule{\textwidth}{0.4pt}

When participants start, they were asked to vote as they did in the tutorial but additional instructions given by the vote buyer 
on the left side of the screen. 
The participants could decide to follow the vote buyer or not. Our idea was to determine whether the participants understood that they can keep their vote while convincing the vote buyer that they follow his choice. Indeed, the dominant strategy for a player, regarding the possibilities offered by the application, is to cast the intended vote while selling a fake tracker to the vote buyer.\footnote{Note that the instructions were formulated without directly revealing this optimal strategy, but the participants should deduce it if they understood the introduction to the study and the explanatory video.}  After computing the tracking number, the participant could choose to send it to the vote buyer or not by clicking on a button. 

\textbf{End of Study} To finish the study, the participants were asked to tell which choice they made - keep their vote intention or follow the vote buyer's instructions - and why. Our last question was about how they felt during the game.

\subsection{Ethical approval}

We obtained ethical approval from our institution's Ethics Panel. 
Our work is compliant with GDPR and the research terms of Prolific.

\subsection{Limitations} \label{method:subsec:limitations}

While having used Prolific brought many advantages, including the reachability of many participants in a small amount of time and good demographics samples, we found some limitations.

First, regarding our trust questionnaire: even though we have built the questionnaire to answer specific needs, we are aware that the questionnaire needs to be tested again to be validated by the community. This first study using it is a first attempt to grasp insights on trust with a specific approach of security perception and acceptance.

Correlations were shown between our measurements (see the next section): some items have been assessed \textit{before} the vote-buying game (trust, usability), while others have been asked \textit{after} the vote buying game (feelings). The correlations found between those measurements could be altered by the game.

Our first pilot study has shown that participants are rushing, likely to increase their reward per hour. Without any guidance, we could not hope that participants will visit all pages in our app, forcing us to make them first test the app through a tutorial rather than exploration.
This is known as ``satisficing bias'' and is acknowledged by Prolific \cite{prolific}. To counter this, we asked the participants to answer questions regarding their understanding in the app: these ``attention checks'' are recommended by Prolific and helped us to lower this bias.

Another limitation concerns our scenario with vote-buying. As for studies in the lab, participants might have a bias to give a good image of themselves, hence answering what would be ethically acceptable \cite{JUSLallemand,Levitt_whatdo}. In this study, some participants justified themselves for having followed the vote buyer because ``this is just a game'', or mentioned their integrity for not having followed him.

Finally, as for other studies, we ask participants to understand new features in a very limited amount of time. More time would be necessary to understand the feature and especially why we implement them.
\section{Results: Evaluation of Voters' Understanding of the receipt-freeness }
\label{understanding:sec:study}
\subsection{Quantitative results}
\subsubsection{Usability and User Experience}

In this section we will explore the results obtained for the user experience and the usability questionnaires. Following to the UX handbook \cite{ueqhandbook}, a result above 0.8 for the UEQ categories would be considered as positive.

We obtained the following results with the UEQ: Attractiveness obtained -0.1 (SD=0.08), Perspecuity obtained -0.41 (SD=0.09), Efficiency obtained 0.31 (SD=0.09), Dependability obtained 0.6 (SD=0.06), Stimulation obtained 0.12  (SD=0.07), Novelty obtained 0.55 (SD=0.07). 

Compared to the previous studies on Selene measuring the user experience through a mobile application \cite{verena,2021marky}, we can see that the web application performed poorly. The attractiveness has been rated as -0.1 (SD=0.08), the usability aspects received the score of 0.16 (SD=0.08) and the hedonic aspects received the score of 0.33 (SD=0.06). At a subscale level, dependability received the higher score with 0.6 (SD=0.06).

Where perspicuity (difficult to learn/easy to learn) was the highest score in \cite{verena} (with 2.16 and 1.90), we obtained the lower score  with -0.41 (SD=0.09). We will discuss the possible reasons in the discussion.

The summary of the SUS results are given in table \ref{uxd:tab:sus}.
We measured effectiveness by asking the participants to give a self assessment of their individual verification step: we asked  if they found their tracking code on the bulletin board. Only 86\% of the participants answered that they found their vote, even though we know that all participants have computed their tracking number.

We measured the efficiency by measuring the time taken by the participants to vote and to compute their tracking code after having logged in to the application. The average time is 57 seconds.

Compared to \cite{2021marky}, again, the web application performed poorly on the satisfaction scale with an average score below 51, considered as ``unacceptable'' in \cite{doi:bangor}. 
We can also note that participants were on average six times faster to vote and verify compared to the lab study in \cite{2021marky}, while the minimum time to cast a vote is almost twelve times faster with the web app, questioning the participants' commitment to the test.

\bgroup
\def\arraystretch{1.5}
\begin{table}[ht]
    \centering
    \footnotesize{}
    \begin{tabular}{c|ccccc|ccccc}
        \multirow{2}{*}{\textbf{Effectiveness}} & \multicolumn{5}{c|}{\textbf{Efficiency}} & \multicolumn{5}{c}{\textbf{Satisfaction}} \\
         & Mean & Median & SD & Min & Max & Mean & Median & SD & Min & Max\\
         \hline
         86 & 57 & 45.5 & 39.65 & 17 & 324 & 48.67 & 45 & 22.81 & 0 & 100
    \end{tabular}
    \caption{Usability results for the web application.}
    \label{uxd:tab:sus}
\end{table}
\egroup
In conclusion, the hypothesis \textbf{H1} is not supported by the experiments: our web app did not provide a positive user experience (scores below 0.8) nor an acceptable usability.

\textbf{Trust} As mentioned above, the questionnaires have been filled after the tutorial phase and before the game. The reason was to let the participants give an evaluation of the app and of its features before we collect the data regarding their understanding. We did not want a specific threat scenario to influence their opinion on the protocol itself.

Overall, trust received an evaluation of $46.81$ ($SD=16.132, Min=4, Max=78$). On the subscale level, the acceptance (over 30) was rated $18.59$ ($SD=7.264, Min=0, Max=30$) and the feeling of security (over 50) was rated $28$ ($SD=20.093, Min=0, Max=48$).

Regarding the grading proposed in section \ref{understanding:sec:trust}, the trust has been evaluated as low by the participants. We can conclude from this result that our hypothesis \textbf{H2} is not supported by our results.

\textbf{Understanding} As a reminder, we evaluate the understanding of the receipt-free 
mechanism as correct when participants kept their vote intention while faking their tracker for the vote-buyer. 
In total, 54 participants have chosen this dominant strategy, i.e. 54 participants understood \textit{correctly} the faking mechanism according to our measurement.

\subsection{Qualitative results}

In App. \ref{app:qualititative} we present the qualitative analysis of the 
the answers from the game and the feedback from our two last questions. Especially we categorise the answer to the question ``Why have you made this choice in the game?'' in terms of the labels money, integrity, understanding, experimenting (wanting to experiment) and miscellaneous. And for the question ``How did you feel during the study'' we use the labels overwhelmed, stressed, offended, good, interested, confident, confused and observed. We relate these labels to the quantitative results below.

\subsection{Relations between variables}\label{sec:rela}

While the questionnaires were filled after the first phase (tutorial) of the user study, the understanding of participants and the qualitative data were collected after the second phase (game). In particular, the vote-buying scenario might have impacted some participants' feedback especially their feeling regarding the study. The following correlations should be considered under this limitation.

\textbf{Trust and Understanding:} When defining trust, our questionnaire was built with the idea that the explanations provided were important to give transparency and to increase the voters' understanding in the application. During the study, we gave explanations through video and text, participants followed a tutorial before playing a game designed to evaluate their understanding of the features. This study design allows us to check the correlation between the Trust results given above and the understanding of voters, measured through their decision in the game.

The understanding has been measured by looking at the capacity of a participant to vote as intended while faking the tracker for the vote buyer. We obtained one group of 54 participants who understood, and another group of 246 participants.
To measure the correlation between trust and understanding, we performed an independent t-test.
The participants who understood the concealing feature gave a statistically higher evaluation of trust ($Mean=51.22$, $SD=15.372$) compared to participants who did not understand it ($Mean=45.84$, $SD=16.163$), $t(298)=2.236$, $p=0.026$. Further, Cohen’s effect size value ($d=0.34$) suggested a small to moderate practical significance.

We conclude the evidence was in favour of  hypothesis \textbf{H3}.

\textbf{Trust and Satisfaction Measures:} We have computed the Pearson correlation coefficient $r=0.561$ ($p=0.01$) 
between our trust and satisfaction measures, meaning that there is a moderate positive correlation between trust and usability.
Similarly, the coefficients 
between Trust and the UEQ's scale are given in table \ref{understanding:tab:pearson}. The values for $r$ are below $0.2$ indicating a weak positive relation.

\vspace{-3mm}
\begin{table}[ht]
    \centering
    \footnotesize{}
    \begin{tabular}{r|cccccc}
         & Attractiveness & Perspicuity & Efficiency & Dependability & Stimulation & Novelty  \\
         \hline
         r & 0.14* & 0.135* & 0.149** & 0.151** & 0.173** & 0.063
    \end{tabular}
    \caption{Correlation coefficients between Trust and UEQ scales, *: $p=0.05$, **: $p=0.01$.}
    \label{understanding:tab:pearson}
\end{table}
\vspace{-8mm}

\textbf{Understanding and Time Spent in the Study:}
When a participant logged in our platform, we timed the session length. 47 participants finished the study in less than 20 min (which was the planned time), whereas the mean was 2155 seconds (35 min and 55 sec). Participants took more time than 
planned, probably because of our attention checks, added after the pilot studies where participants rushed through within five minutes. 
We run a one-way ANOVA test, which has shown no significant difference between the group of participants who understood the game and the others. 

\textbf{Self-explanation/Feeling and Understanding:} Out of the 54 participants who faked their tracking code to send it to the vote buyer, 26 mentioned integrity, 3 money, 17 gave an explanation about their understanding. On the other side, 2 participants explained correctly how the system works, but did not fake their tracking code for the vote buyer. 

Regarding their feeling, 22 participants of the 54 mentioned that they were confused, 25 that they were feeling good, confident or interested in the system, the remaining 7 were feeling observed, stressed, overwhelmed or frustrated.

A Welch ANOVA test between the decision categorization and the understanding shows no significant differences between the five groups ($p>0.05$). Hence, the understanding of participants is not related to the reason for following the vote buyer or not. \\
Similarly, we found no significant differences between the 8 groups of feelings ($p>0.05$). Hence, the understanding of participants is not related by the feeling of participants.

\textbf{Self-explanation/Feeling and Trust:} The relation between the decision's categories and the trust assessments is analyzed with a 1-way ANOVA. The ANOVA test shows a significant difference between the five categories ($F(4,295)=2.872, p=0.023$). A post-hoc Tukey is run to locate differences between categories, and found that participants who mentioned integrity rated trust better (8 points) than those interested in money ($p=0.016$).

On the other side, there was no significant differences between the 8 groups of feelings ($p>0.05$). Hence, the participants' trust (evaluated after the tutorial) was not influenced by their feelings (evaluated after the game).

\textbf{Self-explanation/Feeling and Usability:} We run a 1-way ANOVA test to investigate a relation between the SUS assessments and the self-explanation provided. The test shows a significant difference between the five categories ($F(4,295)=2,729, p=0.029$). A post-hoc Tukey found that participants who mentioned an experimentation gave a better evaluation than those doing the test for money ($p=0.049$).

We also run a 1-way ANOVA test to find a relation between the feeling's categories and the SUS assessments. The ANOVA test shows a significant difference between the 8 categories ($F(7,292)=3.446, p=0.001$). A post-hoc Tukey found that participants who felt interested rated better than those feeling overwhelmed or stressed ($p<0.05$).

The details of the analysis are given in table \ref{understanding:tab:tukeysusfeelings} (we report those with a significant difference only).
\vspace{-3mm}
\begin{table}[ht]
    \centering
    \footnotesize{}
    \begin{tabular}{r|cc}
         & Difference between the means & P value  \\
         \hline
         Experimenting over Money & 15.48 & 0.049\\
         \hline
         Interested over Overwhelmed & 21.34 & 0.039 \\
         Interested over Stressed & 21.34& 0.009 
    \end{tabular}
    \caption{Post-Hoc Tukey significant results between feelings and SUS scores.}
    \label{understanding:tab:tukeysusfeelings}
\end{table}
\vspace{-6mm}
Similarly, we run a 1-way ANOVA to find relations between the UEQ items and the categories for self-explanation and feelings. For self-explanation, no relation was found ($p>0.05$).
We found a relation between the feelings' groups and the UEQ items with statistical significance ($p<0.001$).
Overall, participants having a positive feeling regarding the app rated it better than the other participants with $p<0.05$.

\subsection{Analysis and Discussion}


The big appeal for \textit{moral integrity} was not expected, as our hypothesis was that participants will pick the financially dominant strategy if they understand the features and are rational. The feedback provided by the participants shows that voting is an important matter for them and even if they can deceive a vote-buyer, their own integrity matters more. Unfortunately, we cannot say that our understanding measurement is exhaustive. Furthermore, no relation has been found between the understanding of participants and their self-explanation or feeling regarding the application. However, we have seen that trust and understanding are correlated, which is still satisfying our hypothesis. 

We have seen that the user experience and usability were badly rated. Here we found a moderate correlation between satisfaction and trust, but only a small correlation between UEQ items\footnote{There was no correlation for the item Novelty.} and trust. In the SUS questionnaire, some items concern the acceptance of the tested application, which is one aspect of our trust questionnaire, and might explain the stronger correlation. 
However, we can still argue that a good user interface will benefit a voting application. In \cite{malheirostrustingtolearn} and \cite{kirlappostrust}, the authors mention the signals impacting trust, including usability. We had good results regarding effectiveness and the efficiency, but we failed at convincing the participants that our application was easy to use and enjoyable. 

To explain this, we can look at the feelings that were formulated by the participants. The most expressed feeling was \textit{confusion}: the participants were unsure about the steps to follow in the app. One reason could be the lack of linearity, even if our instructions forced participants to follow a certain workflow in the application. Another reason that was highlighted was the complexity of the study, while Prolific's users are used to surveys, which are linear and require less commitment (in the sense of direct interactions influencing the behaviour of the app) from the user.
The other feelings that were expressed by participants were \textit{stress} and \textit{frustration}: while we tried to provide more guidance to ensure that our app will be correctly tested, it has removed the freedom to navigate and has added complexity to the tasks.

However, we also found that 128 participants had a positive feeling about the study (feeling \textit{good}, \textit{interested}, or \textit{confident}), mentioning their curiosity for online voting or their satisfaction regarding the security of the app. Those participants also rated the usability and UX of the application better than the others, supporting our previous idea of the benefits of a good interface.

We also note that in previous studies using the Selene protocol, for example \cite{verena,2021marky}, the usability and user experience of Selene obtained higher scores. In these studies, Selene was implemented as a mobile app with a linear workflow, and without the faking mechanism. 
As a result, participants just cast their vote and verify that it was correctly recorded. In our study, all participants had to go through the faking feature, which might be the reason why many participants got confused. Further, our application is less linear as participants could navigate through the pages without a unique workflow. 
The lack of linearity and the faking mechanism could have lowered the participants' feedback on usability. 

This low score should be seen in the context of the study
: we wanted to evaluate the full implementation of Selene with all participants testing the faking mechanism. In a real election it is unlikely that all voters need this feature. 
A linear workflow for the vote casting and 
verification phases would probably increase the satisfaction of voters.

Finally, we can hypothesize that the vote buying scenario could have lead to lower the trust in the system: the qualitative feedback received has shown that several people were shocked by the possibility of showing their vote to a coercer/vote buyer, and was sometimes seen 
as vote selling. 
In fact, the mechanism is designed to prevent vote buying, since a vote buyer cannot detect if it is a fake tracker. The security feature and the exacerbation of a possible threat has possibly decreased the trust from the participants, when being misunderstood. We can also note that around 50\% of participants were positive to online voting and more than 90\% did not have negative opinion about it before the study, adding credence to this assumption.

\subsection{Recommendations}\label{sec:rec}

Here we provide a list of recommendations: four concern the development of future voting systems (VS), and two are about the design of user studies (US).

\textbf{[VS] Focus on understandability}
In this study, we have found that participants who understood our security features have rated trust higher than the other participants in general. However, it was highlighted many times that our application was still confusing and tasks were too complex. Also, when providing a new security feature such as a receipt-free 
mechanism, one must ensure that the feature is correctly understood to obtain the desired increase of trust. It remains crucial to provide a transparent interface, with features that are understandable to voters, to increase their trust and acceptance of the voting system.

\textbf{[VS] Provide an easy-to-use interface}
While we must provide understandable and transparent information to participants, it also remains important to keep the interface as simple as possible. People who got stressed and overwhelmed by the application were less satisfied.
Indeed, we found that the participants who rated the application better had a positive feeling during the study. Moreover, we already highlighted that the receipt-free 
feature has added complexity to users who did not have a need for it. 
Hence, we recommend remaining simple and straightforward, keep the workflow as linear and guiding as possible, even though a minimal amount of information must be provided.

\textbf{[VS] Raise awareness and improve education}
Many participants highlighted the illegality of vote buying. 
To them, the fact that the law is already designed to counter some threats is sufficient to trust the system. However, if a voting system is not robust enough nor software independent, it opens a door to attackers who might not be caught. We recommend communicating on good practices in security, on possible risks that could exist in voting and could arise from a misuse of the procedure. Good education, as highlighted in previous work on mental models \cite{zollinger,2021zollinger} and in \cite{kirlappostrust}, is key to trustful applications. 

\textbf{[VS] Adapt the interface to the voters' profile}
We have discussed above that the participants mentioned many times the importance of integrity, and many argued that the scenario was illegal. As we have seen in the previous recommendations, we cannot add complexity or information to the study as it is better to simplify our interface. Furthermore, this also highlights the absence of need for a receipt-free 
feature for most of our participants (in the context of the participants' country). For future implementations, we suggest adapting the interface that will be more realistic to the targeted audience, making receipt-free 
aspects optional. Indeed, participants did not see the necessity of such a feature, as the associated scenario should not happen thanks to the law. Of course, it does not mean that such a scenario is unlikely, especially because we also had positive feedback from people believing it happens, but it will benefit the voters, who will feel more comfortable regarding the protocol they use.

\textbf{[US] Reduce the complexity and simplify (online) user studies}
We have discussed above that many participants were confused during the test, even though instructions were available and guidance was provided during the entire user study. Besides, we know from previous studies \cite{marky2018what,zollinger} that the concept of Verifiability could be hard to understand in the small amount of time that we provided. In addition to Verifiability, we have tested a receipt-free 
feature that has increased the complexity. Further, we have learned that the Prolific's participants needs guidance to follow a study correctly, as they won't take time to explore an application. We recommend simplifying user studies in this context.

\textbf{[US] Use the right tool}
In relation with the different limitations that we have observed with the Prolific platform, we further recommend to design in-person interviews for complex studies about understanding. The bias of satisficing does not help the participants to focus and to take time to understand the features and new concepts that are provided. In this study, we had a small number of participants who clearly understood the features, and we have seen a correlation between their understanding and trust in the system. For an evaluation of the voters' understanding and of the user experience, in-person studies with focus groups and/or interviews will bring better insights.
\section{Conclusion and Future Work} \label{understanding:sec:ccl}

In this paper, we have defined trust in a voting system, and proposed a new questionnaire to assess it. We also designed and conducted a user study where we evaluated the Selene voting system, including its receipt-free 
mechanism. Our application was tested by 300 participants; we evaluated their experience by measuring their understanding through a unique game design, and assessed their trust in the system using the new questionnaire. While the usability and trust factors were rated low in the experiments, we have found a positive relation between trust and understanding. This allowed us to propose a list of recommendations to increase trust and usability in voting applications, as well as improve future user studies. Our recommendations are: 1) Focus on the understandability, 2) Provide an easy-to-use-interface, 3) Raise awareness and improve education, 4) Adapt the scenario to the audience, 5) Reduce the complexity and 6) Use the right tool. The first four apply to any (verifiable) voting system, the two last concern the execution of such trials.
We have also found some limitations, that one should try to mitigate in future studies. This is the 
first user study to investigate a receipt-free feature in the context of vote-buying. 
For future research, it would be interesting to compare the feedback from another country, where our scenario is more common. Also, as a future experiment, we could set up a two-players game where one participant plays the role of a coercer or vote buyer and another plays the role of the voter, to see if the mechanism is better understood by the participants. This will require more understanding from participants: understanding of the system and of the role they must play. A main course of future work is to apply and validate our trust questionnaire for other e-voting protocols and compare to \cite{acemyan2022trust}.

\bibliographystyle{acm}
\bibliography{main}
\appendix
\section{Qualitative results}\label{app:qualititative}

We analysed the answers from the game and also the feedback provided through our two last questions. We categorized the answers into the following categories:
\begin{itemize}
    \item To the question ``Why have you made this \textbf{choice} in the game?'': money, integrity, understanding, experimenting and miscellaneous.
    \item To the question ``How did you \textbf{feel} during the study'': overwhelmed, stressed, offended, good, interested, confident, confused, observed.
\end{itemize}

Then, two researchers independently coded the interviews and compared their findings. They discussed the categorization and solved the disagreements. We tracked the disagreements and Cohen's Kappa was 0.841 for the self-explanation question (almost a perfect agreement) and 0.714 for the feeling question (strong agreement). In the following, we detail our findings.

\subsubsection{Self-explanation}
Regarding the understanding, we measured if a participant managed to fake the tracker while keeping the intended vote, by looking at the decisions recorded on the server. As mentioned above, we counted that 54 of the 300 participants succeeded in doing so. To our first question regarding their choice, 19 participants simply demonstrated their correct \textbf{understanding} of the feature:
    \textit{``I could vote as I wanted and still be paid thanks to the concealed voting.''} (P21),
    \textit{``The game made it perfectly possible for me to select what I personally wanted to vote for, yet fool the vote buyer into thinking I had voted for what they wanted. Win/win situation.''} (P178).

We found that 155 participants mentioned \textbf{moral integrity} as being their motivation behind their decision. Inside this category, participants believe that their intended vote matters, that vote-selling is illegal, or they care about their own integrity and could not disregard their opinion for money:
    \textit{``It's important not to buy votes, even in a game.''} (P10),
    \textit{``There is no point in voting if it is not fair.''} (P48),
    \textit{``I think it is important to vote for yourself and disregard any external influence.''} (P109),
    \textit{``I feel strongly against rigging any form of democratic elections.''} (P175),
    \textit{``I am not followed by greed and I did as I wished.''} (P257).
We can also cite P106 who understood the feature but felt bad about using it: \textit{``I don't like the idea of selling a vote, or of using a feature of the system intended to enhance privacy in order to lie and gain some reward.''}.

Then, we found that 60 participants confessed having followed the vote buyer's instructions because of the \textbf{monetary} incentive, with or without justification:
    \textit{``I knew the situation was not real and I would prefer to receive a higher reward, I would probably not do this in a real situation.''} (P32),
    \textit{``There was an extra incentive to change my vote.''} (P225).

We also found that 22 participants tried to use the application for \textbf{experimenting}, to prove a point, or to see what will happen if they follow the vote buyer, or tried to test a feature. In particular, P5 followed the vote buyer and explained: \textit{``I was intrigued to see what would happen if I selected that option''}.
Other participants wanted to prove the lack of security of the voting protocol: 
    \textit{``I wanted to show how corruptable this method of voting is.''} (P160),
    \textit{``It shows that there are indeed vulnerabilities in the system and that it could be easily influenced.''} (P239).
Some other participants put their trust in the system and wanted to see if the application can deal with corruption, e.g. P273: \textit{``To test the system and how corruption can be detected and avoided''}.

Finally, 44 participants gave a feedback that could not be classified in the above categories, for various reasons: it does not answer our question, e.g. P28 \textit{``That's how I felt''}, or their feedback was unclear, e.g. they might be explaining their voting choice \textit{``It is very important to contain the pandemic''} (P161).

\subsubsection{Feeling}

In this section, we analyse the various emotions reported by the participants. First, 89 participants mentioned their \textbf{confusion} and difficulties to understand the application. Especially, the verification and the concealing feature were hard to understand for participants:
    \textit{``A little confused when it came to the conceal vote part.''} (P12),
    \textit{``I felt a bit confused by the complexity, why is all the verification necessary? Normally when I vote, I vote. And that's it.''} (P171).

Then, 66 participants said that they felt \textbf{good}, enjoyed the study, were calm or relaxed:
    \textit{``Pleased that I was able to conceal my true vote.''} (P137),
    \textit{``During the test I felt calm and in control''} P(267).
In this category, some participants were a bit confused but after the tutorial they felt comfortable, e.g. P70: \textit{``I felt fine, it was slightly confusing first time round.''}.

We counted that 34 participants felt \textbf{confident} or focused during the study, e.g. P91: \textit{``I felt confident in picking my own vote and not changing. Felt safe that my personal information wouldn't be compromised.'',} or P183: \textit{``I felt very positive and very focused''}.

Then, 28 participants were \textbf{interested}, curious or motivated by the study:
    \textit{``Interested to see the potential future of voting.''} (P273),
    \textit{``During the test I found it interesting and creative however it was a bit hard to use.''} (P80).

We counted that 30 participants were annoyed, \textbf{offended} or frustrated by the study's steps, especially the vote-buying scenario, for example:
    \textit{``I felt annoyed by your frustrating voting system.''} (P77),
    \textit{``I felt frustrated that someone could buy peoples vote to get them to vote the way that they wanted.''} (P125),
    \textit{``I felt unsure at points. Offended at the thought of a lack of true democracy.''} (P68).

Then, 19 participants highlighted that the study was \textbf{overwhelming} and tiring, complaining about their lack of understanding or complexity of the tasks again. For example:
    \textit{``During the test I felt exhausted and tired.''} (P14),
    \textit{``I felt overwhelmed by the system''} (P103).

Finally, 6 participants emphasised that they were feeling \textbf{observed} or manipulated, e.g. P113 who said that \textit{``during the test, I felt manipulated''}.

\end{document}